\documentclass[twocolumn,preprintnumbers,amsmath,10pt,aps,pra]{revtex4-1}
\usepackage{graphicx}
\usepackage{color}
\usepackage{hyperref}
\usepackage{subfigure}
\usepackage{graphicx}
\usepackage{type1cm}
\usepackage{eso-pic}
\usepackage{color}
\usepackage{amsmath}
\usepackage{amssymb}
\usepackage{blindtext}

\usepackage{mathptmx}
\def\blue#1{\textcolor{black}{#1}}
\def\bluee#1{\textcolor{black}{#1}}

\def\bluea#1{\textcolor{black}{#1}}

\def\blueq#1{\textcolor{black}{#1}} 
\def\bluek#1{\textcolor{black}{#1}}

\begin{document}



\title{Multiplexing scheme for simplified entanglement-based large-alphabet quantum key distribution}

\affiliation{SUPA, Institute of Photonics and Quantum Sciences, Heriot-Watt University, Edinburgh EH14 4AS, United Kingdom}

\author{Adetunmise Dada}
 \email{a.c.dada@hw.ac.uk}
\affiliation{SUPA, Institute for Photonics and Quantum Sciences, School of Engineering and Physical Sciences, Heriot-Watt University, Edinburgh EH14 1AS, United Kingdom}

\begin{abstract}
We propose a practical quantum cryptographic scheme which combines  high information capacity, such as provided by \blue{high-dimensional quantum entanglement}, with the simplicity of a two-dimensional Clauser-Horne-Shimony-Holt (CHSH) Bell test for security verification. By applying a state combining entanglement in a two-dimensional degree of freedom, such as photon polarization, with \bluee{high-dimensional correlations} in another degree of freedom, such as photon orbital angular momentum (OAM) or path, the scheme
provides a considerably simplified route towards security verification in quantum key distribution (QKD)  aimed at \bluee{exploiting high-dimensional quantum systems} \blue{for increased secure key rates. It also benefits from security against collective attacks 
 and is feasible using currently available technologies}. 
\end{abstract}

\maketitle
\section{Introduction}
Cryptography is one of the most promising applications of quantum science~\cite{PhysRevLett.67.661}. With the recent demonstrations of high-dimensional two-photon entanglement using time bins~\cite{Stucki2005,PhysRevA.73.031801} and OAM~\cite{Dada2011,krenn2014generation}, large-alphabet entanglement-based quantum key distribution (QKD)  systems  become closer to their real-world implementations and applications. 
The traditional approach to large-alphabet QKD based on Bell's theorem involves \blue{encoding a key in a high dimensional degree of freedom, such as photon OAM, and} verifying the security of the generated key using a test of a Bell inequality \blue{which requires  projective measurements in high-dimensional mutually unbiased bases~\cite{PhysRevA.79.052101}}. This is a straighforward generalization of the original protocol introduced by Ekert in 1991 (E91)~\cite{PhysRevLett.67.661,PhysRevA.67.012310} and \blue{its modifications, such as proposed in Ref.~\cite{acin2006efficient}.}

E91-based protocols have been demonstrated for qubits using \blue{polarisation}~\cite{PhysRevA.78.020301} and 
 using time-energy entanglement
 ~\cite{PhysRevLett.84.4737}. A Bell-type test of energy-time entangled qutrits has also been realised~\cite{PhysRevLett.93.010503}. 
Reported \blue{Bell-test-based QKD} experiments using OAM qutrits~\cite{1367-2630-8-5-075} have implemented a randomized selection of dichotomous measurements instead of full projective measurements in a 3-dimensional state space. 
Although projective measurement for detection of high-dimensional OAM states of light with up to 11 different outcomes 
is now within reach~\cite{PhysRevLett.105.153601,2040-8986-13-6-064006}, it still  remains an experimental challenge to perform \blue{them} in arbitrary qudit bases. 
 In the case of high-dimensional time-bin states, such unitary operations would require multi-path interferometric setups which become \blue{too} cumbersome to implement for a high number of dimensions. \blue{Although a scheme for large-alphabet QKD has been proposed and realized using energy-time entanglement}~\cite{PhysRevLett.98.060503}, the applicability of this scheme is specific to this kind of entanglement and the security verification is highly device dependent as it places stringent conditions on timing resolutions of the detectors, which limits \blue{ key generation rates}.   

Security verification of quantum key distribution schemes is a complicated problem in general. Security proofs have been provided for Bell-test-based QKD against the so-called collective attacks~\cite{PhysRevLett.78.2256} as well as the most general coherent attacks in the standard security scenarios~\cite{PhysRevLett.98.230501}. However, proofs of device-independent security against these sophisticated attacks  are not yet available in the case of entangled qudits requiring Bell tests generalised to high-dimensions~\cite{pironio2009device}.

Here, we propose an approach to large-alphabet entanglement-based QKD which circumvents these problems by 
 avoiding the need to perform high-dimensional unitary rotations required for measurements in different mutually unbiased bases, resulting in a much simplified measurement setup. The scheme presented here also benefits from security proofs for QKD based on entangled qubits against collective attacks. 
 \blue{Our approach} is in principle applicable to any system in which it is possible to create bipartite two-dimensional entanglement  in one degree of freedom and high-dimensional  correlations in another. Although 
we will use an example with photon polarization and OAM to illustrate the protocol, the principle can be applied to other systems  
 using other degrees of freedom to encode \blue{large} secret keys.  

\blue{The very essence of large-alphabet QKD is the possibility of a large {\em rate} of key generation. In practice, for a given entanglement-based QKD system,  the minimum applicable coincidence detection time window $\Delta t$  is an important  factor limiting the maximum rate at which it is possible to generate secure keys per run, i.e., a single transmission and detection of the source state. The higher the number of dimensions offered by the source state, the higher the maximum possible key rate per run for a given $\Delta t$}. 
 The development of OAM sorters \blue{makes} genuine large-alphabet key generation using up to 11-dimensional OAM entanglement \blue{feasible}. This will also allow for a higher data rate per photon pair, as the detection of the photonic qudits would not need to be implemented as (probabilistic) dichotomous measurements as has been the case in previous experiments~\cite{1367-2630-8-5-075,Dada2011}. It is also straightforward to implement projective measurements in computational (unrotated) time-bin bases. In what follows, we will first describe the existing generalizations of the E91 protocol. We will then describe the source state, measurement setup, and security considerations for our proposed scheme. Finally, we will conclude with a few remarks on the realizability of the proposed experimental implementations. 

\section{Generalized E91 protocol}
\label{sec:gene91expl}
To establish our scheme,  let us first review the \bluea{basic} entanglement-based large-alphabet QKD \blue{resulting from a direct generalization of the E91 protocol and its variants}. Assume a source producing photon pairs in the state 
\begin{equation}
\label{eqn:sxasffw}
|\Phi\rangle =\tfrac{1}{\sqrt{d}} \sum^{d-1}_{j=0} {\left| j \right\rangle _A \otimes \left| {j} \right\rangle _B }.
\end{equation} 
Here we use the notation $\left| { x ,y} \right\rangle\equiv \left| { x} \right\rangle\otimes\left| { y} \right\rangle$, where $\otimes$ denotes tensor product.
In terms of OAM eigenstates $|\ell\rangle$ for example, this may be written as the maximally entangled state
\begin{equation}
\label{eqn:sxasffw2}
|\Phi\rangle =\tfrac{1}{\sqrt{d}} \sum^{\ell=+[d/2]}_{\ell=-[d/2]}h(\ell){\left| \ell \right\rangle _A \otimes \left| {-\ell} \right\rangle _B },
\end{equation} 
where $h(\ell)=1$ for all $\ell$ when $d$ is odd, and $h(\ell\ne0)=1$, \blue{$h(0)=0$} when $d$ is even.

In a Bell inequality test experiment, each of the communicating parties `Alice' (A) and `Bob' (B) \blue{will} have a photon OAM detector with  \blue{$D$ outcomes per setting} and two settings/measurements: $\{A_1,~A_2\}$ and $\{B_1,~B_2\}$ respectively, which maximize 
 Bell inequality violation. \blue{ For the  QKD scheme, there is an additional setting for each detector, i.e., $A_3$ for Alice and $B_3$ for Bob, chosen to produce perfect correlations. In a variant of Ekert's scheme modified for increased key generation efficiency~\cite{acin2006efficient,PhysRevLett.98.230501,pironio2009device}, only Alice's detector uses an additional setting, i.e., $A_0$, which is chosen to produce perfect correlations when Bob measures with setting $B_1$ for the purpose of key generation. Although our scheme is directly applicable to this higher-efficiency version, we mainly illustrate here using Ekert's scheme for clarity.} 
 \begin{figure*}[ht!]
\centerline{\includegraphics[width=1.0\textwidth]{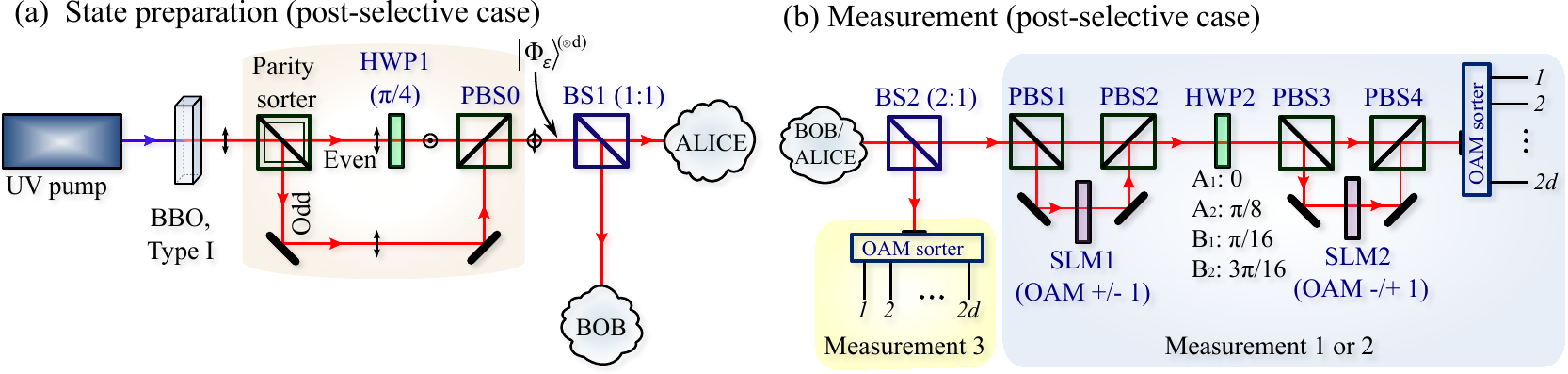}}
\caption{\blueq{(Color online) }Schematic diagram for a suggested implementation of the proposed simplified large-alphabet entanglement-based quantum key distribution using OAM and polarization. (a) Preparation of the \blue{two-photon state $|\Phi_\epsilon\rangle$ [Eq.~\eqref{eq:mstatepoloam}] or $|\Phi^d_\epsilon\rangle$ [Eq.~\eqref{eq:mstatemulti3}] using spontaneous parametric down conversion (SPDC) in a $\beta$-barium borate (BBO) nonlinear crystal cut for type-I spontaneous parametric down conversion.  The preparation uses an OAM parity sorter~\cite{PhysRevLett.88.257901}, a polarizing beam splitter (PBS) and a non-polarising 1:1 beam splitter (BS). (b) Measurement setup for Alice (Bob). Measurements 1, 2, and 3 i.e., $A_j$ and $B_k$ ($j,k=1,2,3$) are respectively selected randomly (e.g., using beam splitters) on Alice's and Bob's side.  For the Bell test, Alice (Bob)  adds (subtracts) $\hbar$ of OAM for vertically polarised photons using PBS1, SLM1 (OAM$\pm1$) and PBS2. Then Alice (Bob) sets the half-wave plate (HWP2) orientation angle to implement the randomly chosen measurement ($A_1$/$B_1$ or $A_2$/$B_2$). HWP2 and PBS3 are used for polarization analysis in the Bell-test, after which Alice (Bob) may then choose to reverse the first operation by using SLM2 (OAM$\mp 1$) and PBS4. OAM sorting~\cite{PhysRevLett.105.153601} is used to resolve the qubit subspaces and/or establish the key (measurement 3).}
}
\label{fig:setup1}
\end{figure*}
Alice and Bob independently choose their settings at random and also note their detection results independently. After sufficiently many measurement runs, Alice and Bob  \blue{perform basis reconciliation through one-way classical post processing~\cite{PhysRevLett.95.080501}, followed by privacy amplification on the raw key.} 

\blue{ 
When the combination $\{A_3,  B_3\}$ (or $\{A_0, B_1\}$) 
 is selected by Alice and Bob,  the measurement results are used for the secret key as they are perfectly correlated on both sides.}  To determine the security of this key, the correlation in the rest of the data will be checked for eavesdropping through a Bell inequality test, for example using Bell inequalities generalised to $d$-outcomes per measurement proposed by Collins {\it et al.}~\cite{PhysRevLett.88.040404}, equivalent to the CHSH-Bell inequality~\cite{PhysRevLett.23.880} \blue{when} $d=2$. Only cases in which the combination of measurement settings involve $A_{1,2}$ and $B_{1,2}$ are used for this test, while  the remaining results are discarded. After basis reconciliation, Bob announces his data for the Bell inequality check, and
Alice computes the value of the Bell parameter $S$. If $S> 2$, then the key is secure and the eavesdropper, Eve, will
not have gained any useful information on the key. The secret key can then be used in any cryptographic communication between Alice and Bob. 

Implementing the above requires full projective measurements in the OAM state basis $\{|\ell\rangle\}$ in a $d$-dimensional subspace, corresponding to  $A_3,B_3$. This \bluea{may} be realized, e.g. for up to $d=11$ using OAM mode sorters as mentioned above. However, full projective measurements whose operators have eigenstates which are OAM superpositions are also required. 
It is nontrivial to realize such measurements because it requires a unitary operation within the high-dimensional OAM subspace being considered before the OAM detection. The implementations of such operations are difficult to derive in general, and have not \blue{yet been} realized experimentally.

\section{Proposed scheme}

\subsection{State preparation}
We propose a state which replaces the need for measurements in high-dimensional rotated bases with the simplicity of a two-dimensional CHSH Bell test for the verification of the security of generated key.
 To appreciate how our source state relates to hybrid entangled states, consider the state expressed in terms of the composite OAM and polarisation \blue{basis} states $|\ell, P\rangle$ (where $\ell$ denotes the OAM, $\ell=-\infty,\ldots,+\infty$; and $P$ denotes the prolarization $P = H,V$) as
\small
\begin{align}
\label{eq:mstatepoloam}
|\Phi_{\epsilon}\rangle=\frac{1}{{\sqrt{ 2d} }}\sum^{n={\rm +}[{d}/{2}]}_{n={\rm -}[{d}/{2}]} &{{{\left| {2n ,H} \right\rangle }_A}{{\left| { {\rm -} 2n ,H} \right\rangle }_B} }  \nonumber\\   &{{\rm +} {{\left| {2n {\rm -} 1,V} \right\rangle }_A}{{\left| { {\rm -} 2n  {\rm +} 1,V} \right\rangle }_B}},
\end{align} 
\normalsize
with $n\ne0$ for even $d$. Note that this state combines \blue{$2d$-dimensional} orbital angular momentum entanglement and  polarization entanglement in a way similar but quite different from the cases of the so-called hyper-entangled~\cite{PhysRevLett.95.260501}, hypoentangled~\footnote{N. K. Langford, Ph.D. thesis, Univ. of Queensland (2007)} or entangled entangled~\cite{PhysRevLett.97.020501} states. In hyper-entanglement, a measurement of OAM will not destroy polarization entanglement and vice versa. In hypoentanglement, measuring either polarization or OAM destroys entanglement in the other degree of freedom. Here measuring OAM completely destroys polarization entanglement, but the converse is not true. 
We note that the division of the subspaces (e.g. into odd and even OAM parities in this example) can also be done in other ways, depending on the specific realization and experimental convenience.  
State~\eqref{eq:mstatepoloam} can be rewritten,  as
\small
\begin{align}
\label{eq:mstatepoloam2}
|\Phi_{\epsilon}\rangle=\frac{1}{{\sqrt{d} }}\sum^{n={\rm }[{d}/{2}]}_{n={\rm-}[d/2]} |\phi\rangle_n,  ~~n\ne0 {\rm~for~even}~d.
\end{align}
\normalsize
Here $|\phi\rangle_{n}$ is an entangled state within the $n$th OAM subspace. Although this state is (hypo)entangled in both polarization and OAM,  only the classical correlation in OAM is strictly necessary for our scheme.
\\
\\
{\bf Source state:} Our source state is of the form
 \begin{align}
\label{eq:mstatemulti2}
|\Phi_{s\epsilon}^d\rangle^{\rm P/D}=\bigotimes_{n={\rm -}[{d}/{2}]}^{n={\rm }[{d}/{2}]}\,|\phi\rangle_{n}^{\rm P/D}, 
\end{align} 
Where $|\phi\rangle_{n}^{\rm P/D}$ is an entangled state in polarisation within an OAM subspace specified by $n$.
The source state could be obtained either by post-selection or deterministically (denoted by superscripts P and D respectively) as outlined below.

\subsubsection{State preparation: Post-selective case}
\label{sec:nondet}
Suppose we define
\begin{align}
\label{eq:mstatepoloam23}
|\phi\rangle_{n}^{\rm P} &= \left(\left| { H_A,H_B} \right\rangle_n + \left| { V_A,V_B} \right\rangle_n\right)/\sqrt{2}, {~\rm where}\\
\label{eq:basdef2}
\left| { H_A,H_B} \right\rangle_n &=  {{{\left| {2n ,H} \right\rangle }_A}\otimes{{\left| { {\rm -} 2n ,H} \right\rangle }_B} } \nonumber\\
\left| { H_A,V_B} \right\rangle_n &=  {{{\left| {2n ,H} \right\rangle }_A}\otimes{{\left| { {\rm -} 2n+1 ,V} \right\rangle }_B} }  \nonumber\\
\left| { V_A,H_B} \right\rangle_n &=  {{{\left| {2n-1 ,V} \right\rangle }_A}\otimes{{\left| { {\rm -} 2n ,H} \right\rangle }_B} }\nonumber\\
\left| { V_A,V_B} \right\rangle_n &=  {{{\left| {2n-1 ,V} \right\rangle }_A}\otimes{{\left| { {\rm -} 2n +1,V} \right\rangle }_B}}.
\end{align}

Note that this state is a combination of $d$ photon pairs, with each pair hypoentangled in both polarisation and OAM in unique OAM subspaces.

A source state for our scheme [of the form Eq.~\eqref{eq:mstatemulti2}] could be obtained by post-selection from
\begin{align}
\label{eq:mstatemulti3}
|\Phi_{\epsilon}^d\rangle=|\Phi_{\epsilon}\rangle^{\otimes d},
\end{align} 
which is a product state of $d$ pairs of OAM-entangled photons where $|\Phi_{\epsilon}\rangle$ is the two-photon state expressed in Eq.~\eqref{eq:mstatepoloam}. 
A proposed scheme to obtain $|\Phi_{\epsilon}\rangle^{\otimes d}$  from common spontaneous parametric down conversion (SPDC) sources is illustrated in Fig.~\ref{fig:setup1}(a). This involves generating OAM entanglement by type-I collinear parametric downconversion with a defined polarization, say horizontal ($|H\rangle$). The co-propagating photon pairs entangled in OAM are passed through an OAM parity (even/odd) sorter~\cite{PhysRevLett.88.257901}. A half-wave plate is then inserted after one of the output arms which rotates $|H\rangle$ to vertical polarization $|V\rangle$, coupling OAM parity to polarization.  The state represented in Eq.~\eqref{eq:mstatemulti3} could then be generated by choosing parameters of the SPDC source to create more than one entangled photon pair simultaneously. It is well known that a desired probability of multiple pair generation per pump pulse can be achieved according to the theoretical $d$-pair creation probability~\cite{PhysRevA.61.042304} 
\begin{align}
\label{eq:multiprobd}
p_d=(d+1) \text{sech}^4(\tau) \tan ^{2 d}(\tau),
\end{align}
where $\tau$ is a real-valued coupling coefficient which is proportional to the product of the pump amplitude and the coupling constant
between the electromagnetic field and the nonlinear crystal.
The source state for our scheme can then be obtained by final postselection on state represented by Eq.~\eqref{eq:mstatepoloam}. This can be done by registering only the values of $n$ for which both Alice and Bob have a \bluee{single} detection \bluee{each} per OAM subspace in one run. To achieve this, it suffices to use detectors which distinguish between  zero,  one, and more than one photon.   Such detectors have been experimentally demonstrated~\cite{Kwiat:94,Takeuchi1999}. Also, actual photon-number-resolving detectors have been realised (e.g., see~\cite{kardynal2008avalanche,dauler2009photon}) with increased detection efficiencies~\cite{calkins2013high}.

\subsubsection{State preparation: Deterministic case}
\label{sec:detprepst1}
\begin{figure*}[t!]
\centerline{\includegraphics[width=0.95\textwidth]{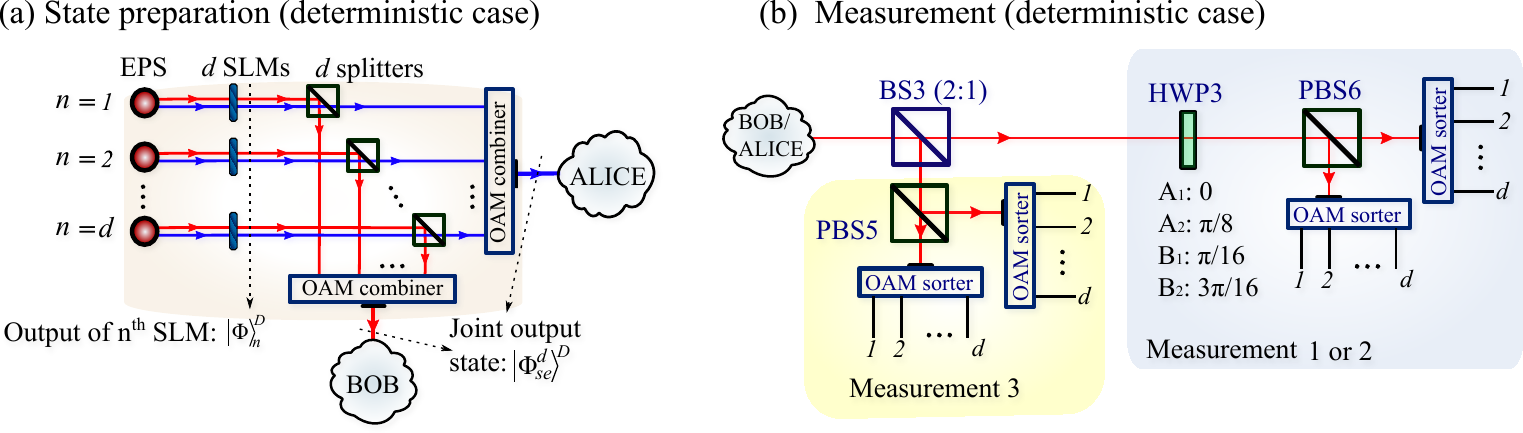}}
\caption{\blueq{(Color online) }Schematic diagram for a deterministic implementation of the proposed QKD scheme (a) Suggested preparation of the source state using an array of single-entangled-photon-pair sources (EPS), spatial light modulators (SLMs), splitters, and OAM combiners. (b) Suggested measurement setup for Alice (Bob). Measurements 1, 2, and 3, i.e., $A_j$ and $B_k$ ($j,k=1,2,3$) are respectively selected randomly (e.g., using beam splitters) on Alice's and Bob's side.  For the Bell test, Alice (Bob) sets the half-wave plate (HWP3) orientation angle to implement the randomly chosen measurement. HWP3 and PBS6 are used for polarization analysis in the Bell-test. As in Fig.~\ref{fig:setup1},  OAM sorting is used to resolve the qubit subspaces and/or establish the key.}
\label{fig:setup2}
\end{figure*}

A more suitable approach, however, is to prepare the source state in a deterministic way by, e.g., using an array of $d$ polarization-entangled-photon sources (EPS) generating exactly one photon pair at a time. Existing semiconductor quantum dot (QD) systems provide a suitable platform for single photon generation~\cite{PhysRevLett.96.130501,nature04446}, as well as generation of entangled photon pairs on demand with high efficiency~\cite{nphoton.2013.377}. Rapid experimental progress is also  being made towards implementing arrays consisting of several QD high-fidelity-entangled-photon-pair  emitters on the same chip~\cite{Juskanphoton2013}. Here, we propose a setup utilising $d$ EPS (see Fig.~\ref{fig:setup2}).  OAM of $\ell=n$, for example,  is then individually imprinted on photons emitted by the $n^{\rm th}$ source to yield state \bluek{$|\phi\rangle_n$,} resulting in  basis states that are assigned as shown below. The photons in an entangled pair are usually generated using the biexciton-exciton-vacuum cascade, and are separable based on their wavelength.   
Individual photons from different pairs may then be combined into one beam using an OAM combiner (i.e., a coherent OAM sorter operated in reverse) to obtain the source state $|\Phi_{s\epsilon}^d\rangle$ [Eq.~\eqref{eq:mstatemulti2}]. 
For this case, the entangled state $|\phi\rangle_n^{\rm D}$ within the $n$th subspace defined as
\begin{align}
\label{eq:QDmstatepoloam}
|\phi\rangle_{n}^{\rm D} &= \left(\left| { H_A,H_B} \right\rangle'_n + \left| { V_A,V_B}\right\rangle'_n\right)/\sqrt{2}, {~\rm where}\\
\label{eq:basdef23}
\left| { H_A,H_B}\right\rangle'_n &=  {{{\left| {n ,H} \right\rangle }_A}\otimes{{\left| { {\rm } n ,H} \right\rangle }_B} } \nonumber\\
\left| { H_A,V_B} \right\rangle'_n &=  {{{\left| {n ,H} \right\rangle }_A}\otimes{{\left| { {\rm } n ,V} \right\rangle }_B} }  \nonumber\\
\left| { V_A,H_B} \right\rangle'_n &=  {{{\left| {n ,V} \right\rangle }_A}\otimes{{\left| { {\rm } n ,H} \right\rangle }_B} }\nonumber\\
\left| { V_A,V_B} \right\rangle'_n &=  {{{\left| {n ,V} \right\rangle }_A}\otimes{{\left| { {\rm } n,V} \right\rangle }_B} }.
\end{align}

We note that the source state[Eq.~\ref{eq:mstatemulti2}] is essentially the same for both the probabilistic and deterministic preparations except for a change in the basis state assignment of the OAM measurement. This basis selection is simply for the convenience of experimental implementation specific to each method of state preparation. 

\subsection{\em Measurement settings} \blue{As in the standard case for the generalised E91 protocol described in Section~\ref{sec:gene91expl}}, our scheme using the state \eqref{eq:mstatemulti2} also involves three measurement settings randomly and independently chosen by Alice and Bob. However, the settings $A_{1,2}$ and $B_{1,2}$ are now \blue{achieved using} polarization measurements for maximal CHSH-Bell inequality violation. These measurement settings each have two outcomes ``+'' and ``-''. For key generation, $A_3$ and $B_3$,  (\bluee{or} $A_0$ and $B_1$)  are the same as described above. An important aspect of our scheme is to perform both key generation and Bell tests {\em individually} in each $n$th subspace (or channel),  and {\em simultaneously} for all $n=1,...,d$, using the same Bell-test setup.  
\subsubsection{Measurement: Post-selective case}
To achieve the simultaneous measurements for the case of the non-deterministic state preparation outlined in Section~\ref{sec:nondet} above, Alice \bluee{and} Bob need to first perform local operations which make the respective OAM states degenerate for orthogonal polarisations of Alice's and Bob's photons within each $n$th subspace, i.e., to disentangle the polarisation and OAM degrees of freedom. 
 This can be achieved if, e.g., Alice (Bob) subtracts (adds)$\hbar$ of OAM for the vertically polarised photons 
[using the combination of PBS1, SLM1 and PBS2 in Fig.~\ref{fig:setup1} (b)].  This operation by Alice and Bob can be described by the transformations $\hat{Q}_A$ and $\hat{Q}_B$ where
\begin{eqnarray}
\label{eq:opr1}
\hat{Q}_A&=&\sum^d_{n=1}\,{{{\left| {2n ,H} \right\rangle }_A}} {{{\left\langle {2n ,H} \right| }_A}} + {{{\left| {2n ,V} \right\rangle }_A}} {{{\left\langle {2n - 1 ,V} \right| }_A}}\\
\hat{Q}_B&=&\sum^d_{n=1}\,{{{\left| {-2n ,H} \right\rangle }_B}} {{{\left\langle {-2n ,H} \right| }_B}} + {{{\left| {-2n ,V} \right\rangle }_B}} {{{\left\langle {-2n - 1 ,V} \right| }_B}}.\nonumber
\end{eqnarray}
 Note that this only causes a transformation of the basis states defined in Eq.~\eqref{eq:basdef2} as follows,  
  \begin{align}
\label{eq:basdef3}
\left| { H_A,H_B} \right\rangle_n &\xrightarrow{\hat{Q}_A\otimes\hat{Q}_B} \left| { H_A,H_B}\right\rangle^Q_n=  {{{\left| {2n ,H} \right\rangle }_A}\otimes{{\left| { {\rm -} 2n ,H} \right\rangle }_B} }, \nonumber\\
\left| { H_A,V_B} \right\rangle_n &\xrightarrow{\hat{Q}_A\otimes\hat{Q}_B} \left| { H_A,V_B} \right\rangle^Q_n = {{{\left| {2n ,H} \right\rangle }_A}\otimes{{\left| { {\rm -} 2n ,V} \right\rangle }_B} },  \nonumber\\
\left| { V_A,H_B} \right\rangle_n &\xrightarrow{\hat{Q}_A\otimes\hat{Q}_B}\left| { V_A,H_B} \right\rangle^Q_n =  {{{\left| {2n ,V} \right\rangle }_A}\otimes{{\left| { {\rm -} 2n ,H} \right\rangle }_B} },\nonumber\\
\left| { V_A,V_B} \right\rangle_n &\xrightarrow{\hat{Q}_A\otimes\hat{Q}_B}  \left| { V_A,V_B} \right\rangle^Q_n= {{{\left| {2n,V} \right\rangle }_A}\otimes{{\left| { {\rm -} 2n ,V} \right\rangle }_B} }.
\end{align}
A combination of a HWP and a PBS can now carry out the Bell-test polarisation measurements ($A_{1,2}$ or $B_{1,2}$) for each value of $n$.

\blue{We can write the CHSH inequality in the $n$th subspace as}
\begin{equation}
\label{eq:chsh}
S_n= E_n(A_1,B_1)-E_n(A_1,B_2)+E_n(A_2,B_1)+E_n(A_2,B_2)\le2,
\end{equation}
where the correlation coefficients of the measurement $A_i$ performed by Alice and $B_j$ by Bob are defined as
\begin{equation}
\label{eq:chsh1}
E_n(A_i,B_j)={P_n(A_i=B_j)-P_n(A_i\ne B_j)}.
\end{equation}
$P_n(A_i=B_j) $ and $P_n(A_i\ne B_j) $ are probabilities for equal and unequal outcomes respectively, determined experimentally using the coincidence \bluee{rates within} each $n$th subspace.
The detector settings for the CHSH Bell inequality violation could be specified as measurements in the bases 
$\{|m_+(\theta)\rangle,|m_-(\theta)\rangle\}$, where \blue{
\begin{align}
\label{eq:basez}
|m_+(\theta)\rangle = &-\cos(2\theta) {{\left| { { \pm} 2n ,H} \right\rangle }} +\sin(2\theta) {{\left| { { \pm} 2n ,V} \right\rangle }}, \nonumber\\
|m_-(\theta)\rangle = &\sin(2\theta){{\left| { { \pm} 2n ,H} \right\rangle }}  + \cos(2\theta) {{\left| { { \pm} 2n ,V} \right\rangle }} .
\end{align} 
 In the $\pm$ sign above, `$+$' applies to Alice and `$-$' applies to Bob. A half-wave plate oriented at an angle $\theta$ rotates the measurement basis of a polarizing beam splitter (PBS) i.e., $\{|H\rangle,|V\rangle\}$ to $\{|m_+(\theta)\rangle,|m_-(\theta)\rangle\}$.
 }
If we set 
\begin{equation}
\label{eq:msettings}
\theta_1^a=0,~\theta_2^a=\pi/8,~\theta_1^b=\pi/16,~{\rm and }~\theta_2^b=3\pi/16
\end{equation} as values of $\theta$ for $A_1, A_2, B_1$ and $B_2$ respectively \blueq{so that Alice and Bob always measure in bases which are mutually unbiased with respect to each other}, then we will \blueq{ensure the commutativity of Alice's and Bob's measurement outcomes} and get the maximal violation of $2\sqrt{2}$ for each $n$th subspace of state~\eqref{eq:mstatemulti2}. 
\blue{Using the basis notation 
defined above [Eq.~\eqref{eq:basdef3}], the corresponding Bell operator~\cite{PhysRevA.51.R1727,PhysRevLett.68.3259,PhysRevA.65.052325} can be written 
as
\blueq{
\begin{align}
\label{eq:belloperatorn}
\hat{S}_n = &\bluek{\sqrt{2}(}\left|  H_A,H_B\right\rangle^Q_n \left\langle { H_A,H_B} \right|^Q_n  + \left| { V_A,V_B} \right\rangle^Q_n \left\langle { V_A,V_B} \right|^Q_n\nonumber\\
 &+\left| { H_A,H_B} \right\rangle^Q_n \left\langle { V_A,V_B} \right|^Q_n  + \left| { V_A,V_B} \right\rangle^Q_n \left\langle { H_A,H_B} \right|^Q_n\nonumber\\
 &-\left| { H_A,V_B} \right\rangle^Q_n \left\langle { H_A,V_B} \right|^Q_n  - \left| { V_A,H_B} \right\rangle^Q_n \left\langle { V_A,H_B} \right|^Q_n\nonumber\\
 &+\left| { H_A,V_B} \right\rangle^Q_n \left\langle { V_A,H_B} \right|^Q_n  + \left| { V_A,H_B} \right\rangle^Q_n \left\langle { H_A,V_B} \right|^Q_n\bluek{)}.
\end{align}
}
Obtaining the statistical data for the Bell test requires either carrying out a detection which resolves both polarisation and OAM or, as illustrated in Figure~\ref{fig:setup1} (b), reversing operation $\hat{Q}_{A/B}$ to re-establish OAM-polarisation entanglement (using SLM2 and PBS4), and then carrying out OAM detection. 
We define the operations to reverse $\hat{Q}_{A/B}$ as 
\begin{eqnarray}
\label{eq:oprev1}
\hat{Q}_A^-&=&\sum^d_{n=1}\,{{{\left| {2n ,H} \right\rangle }_A}} {{{\left\langle {2n ,H} \right| }_A}} + {{{\left| {2n ,V} \right\rangle }_A}} {{{\left\langle {2n + 1 ,V} \right| }_A}}\\
\hat{Q}_B^-&=&\sum^d_{n=1}\,{{{\left| {-2n ,H} \right\rangle }_B}} {{{\left\langle {-2n ,H} \right| }_B}} + {{{\left| {-2n ,V} \right\rangle }_B}} {{{\left\langle {-2n + 1 ,V} \right| }_B}}.\nonumber
\end{eqnarray}
Since the state within the $n$th subspace [Eq.~\eqref{eq:mstatepoloam23}] is maximally entangled, it gives a maximal violation of the CHSH inequality based on operator~\eqref{eq:belloperatorn}
\begin{align}
\label{eq:troam2s}
 S_n(\left|  \phi\right\rangle_n \left\langle  \phi \right|_n )={\rm Tr}(\hat{S}_n  \left|  \phi\right\rangle_n \left\langle  \phi \right|_n )=2\sqrt{2}\ge 2. 
\end{align}
}

\subsubsection{Measurement: Deterministic case}
When the state is prepared deterministically as described in Section~\ref{sec:detprepst1}, operators $\hat{Q}_A^{(-)}$ and $\hat{Q}_B^{(-)}$ [Eqs.~\eqref{eq:opr1} and \eqref{eq:oprev1}] are not necessary for the measurements. As in the non-deterministic case, the Bell test  is carried out using a combination of a HWP and PBS [see Fig.~\ref{fig:setup2} (b)], but photon number resolution and final postselection are not required. Due to the difference in basis assignment in this case, we redefine the detector settings for the CHSH Bell inequality violation as measurements in the bases 
$\{|m_+(\theta)\rangle',|m_-(\theta)\rangle'\}$, where 
\begin{align}
\label{eq:basezdet}
|m_+(\theta)\rangle' = &-\cos(2\theta) {{\left| { n ,H} \right\rangle }} +\sin(2\theta) {{\left| {  n ,V} \right\rangle }}, \nonumber\\
|m_-(\theta)\rangle' = &\sin(2\theta){{\left| {  n ,H} \right\rangle }}  + \cos(2\theta) {{\left| {  n ,V} \right\rangle }}.
\end{align} 
The optimum settings (specified by $\theta$) for the HWP are the same as in Eq.~\eqref{eq:msettings} above, and the resulting Bell operator for this case [see Eq.~\eqref{eq:basdef23}] is
\begin{align}
\label{eq:operatorn2}
\hat{S}_n = &\bluek{\sqrt{2}(}\left|  H_A,H_B\right\rangle'_n \left\langle { H_A,H_B} \right|'_n  + \left| { V_A,V_B} \right\rangle'_n \left\langle { V_A,V_B} \right|'_n\nonumber\\
 &+\left| { H_A,H_B} \right\rangle'_n \left\langle { V_A,V_B} \right|'_n  + \left| { V_A,V_B} \right\rangle'_n \left\langle { H_A,H_B} \right|'_n\nonumber\\
 &-\left| { H_A,V_B} \right\rangle'_n \left\langle { H_A,V_B} \right|'_n  - \left| { V_A,H_B} \right\rangle'_n \left\langle { V_A,H_B} \right|'_n\nonumber\\
 &+\left| { H_A,V_B} \right\rangle'_n \left\langle { V_A,H_B} \right|'_n  + \left| { V_A,H_B} \right\rangle'_n \left\langle { H_A,V_B} \right|'_n\bluek{)}.
\end{align}

The state represented by Eq.~\eqref{eq:QDmstatepoloam} is also maximally entangled within the $n$th subspace for this case, and it gives a maximal violation of the CHSH inequality based on operator~\eqref{eq:operatorn2} when the key has not been eavesdropped. 

\subsection{ Security against collective attacks}  
\begin{figure}
\centerline{\includegraphics[width=0.4\textwidth]{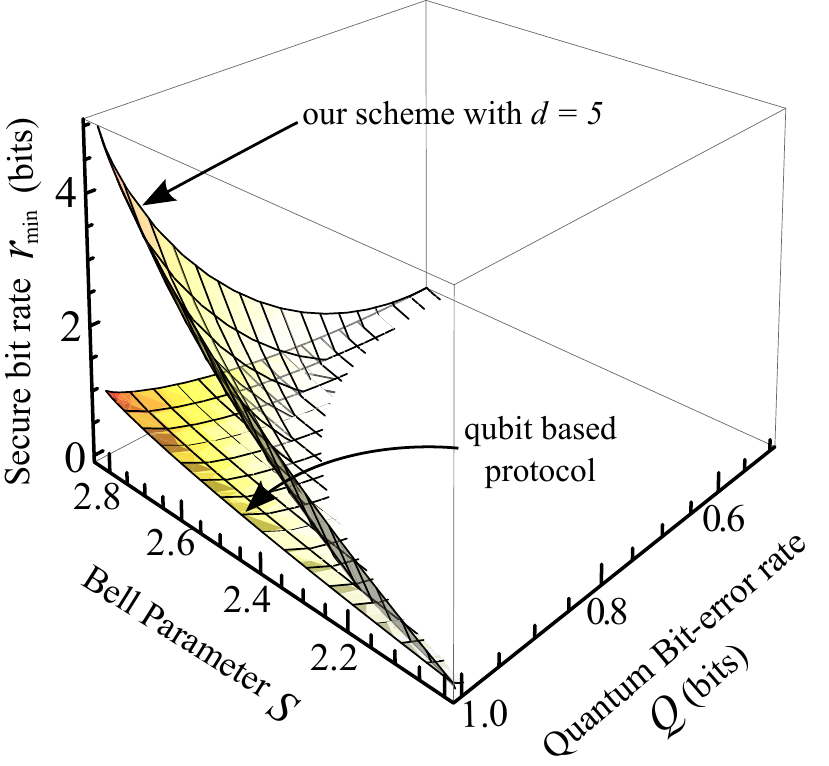}}
\caption{\blueq{(Color online) }Comparison of the minimum secure key rate $r_{min}$ as a function of the Bell parameter $S$ and the quantum bit-error rate (QBER) $Q$, in a single run, for our scheme with a qubit-based E91-type protocol~\cite{PhysRevLett.98.230501,pironio2009device}. We assume that the quantum bit error rate $Q$ and Bell parameter for each channel is the same, i.e., $Q_n=Q$, and  $S_n=S$ respectively for all $n$. Our scheme shows a $d$-fold enhancement in secure key rate.}
\label{fig:securerate1nd}
\end{figure}

 Any eavesdropping of the key is essentially a measurement strategy that will destroy polarisation entanglement which is used to establish the key. This in turn degrades the CHSH Bell inequality violation~\cite{PhysRevLett.67.661}  \blue{in any respective OAM subspaces}.  
   A {\em collective attack} is one in which the eavesdropper (Eve) applies the same operation on each of Alice's and Bob's particles, \bluea{but has no} other limitations.  In particular,  
she  is allowed to have access to a string of qubits from Alice/Bob at one time, \bluee{and to other dimensions of their particle states, even possibly unknown to Alice/Bob}. \bluee{Since Eq.~\eqref{eq:mstatemulti2} is a product state of $d$ entangled qubits pairs,} our scheme is essentially a multiplexing of multiple polarisation-entangled qubit pairs by means of a higher-dimensional degree of freedom, followed by independently testing the CHSH Bell inequality simultaneously---Eve's access to one or more source states in our scheme is equivalent to her access to a string of qubits on which she can perform joint (coherent) measurements. Therefore, the security of our scheme is completely guaranteed by the security of the individual qubit-based schemes against collective attacks~\cite{PhysRevLett.78.2256}. 
This, in turn, implies security against the most general, so-called coherent attacks~\cite{PhysRevLett.78.2256, PhysRevLett.79.4034} \blueq{if  an application of the exponential quantum {\em de Finetti} theorem can be made~\cite{renner2007symmetry}.  This is indeed the case in our scheme (under the assumption of finite-dimensional subsystems) because our source state is invariant under permutation of Alice and Bob, and their measurement outcomes are commutative, as mentioned above [Eq.~\eqref{eq:msettings}]}.  
These  results apply fully to our large-alphabet protocol since it is equivalent to simultaneous but independent 2-qubit secure protocols.
  The \bluee{total} bit rate generated securely against collective attacks as a function of Bell parameters $S_n$ 
 \bluee{ can therefore be written as}~\cite{PhysRevLett.98.230501}
\begin{align}
\label{eq:security-rate}
r\ge\sum^{}_{n}\, 1-h(Q_n)-h\left( \frac{1+\sqrt{(S_n/2)^2-1}}{2}\right)
\end{align} 
where $h$ is the binary entropy and $Q_n$ is the quantum bit error rate for channel $n$. As shown in Fig.~\ref{fig:securerate1nd}, the larger the measured Bell violation, the higher the secure key rate per run.  \bluee{Our scheme gives} a $d$-fold enhancement over the traditional 2-qubit schemes as a large-alphabet scheme, but uses a much simplified Bell-test measurement setup compared to traditional large-alphabet \bluek{schemes}. 
   
\blue{The implications of loopholes for QKD based on Bell's theorem is worthy of some mention here. Closing the locality loophole in general requires enforcing a space-like separation between Alice and Bob as required for testing non-locality~\cite{PhysRevLett.49.1804}, but  in the context of our QKD scheme, it would be sufficient to guarantee that no quantum signals 
  can travel from Alice to Bob by ensuring proper isolation of Alice's and Bob's locations~\cite{pironio2009device}. Also, a proper closure of the detection loophole is required for completely guaranteed security. This seems promising as it has already been achieved  in a photon-based Bell-test experiment~\cite{giustina2013bell}.}
\section{Conclusion }
\blue{Our scheme offers significant advantages over current generalised E91 schemes.} It results in a greatly simplified security verification and key generation setup which does not get more complicated with increasing $d$, except for an increase in the  number of output ports of the OAM sorting device.  It thereby provides a route to boosting the secure key rates in entanglement-based QKD without the usual increased complexity of Bell tests in high dimensions. 
It also benefits from the relative tolerance two-dimensional Bell tests to  measurement error. Although it is known that the amount of violation for an actual $D$-dimensional Bell test increases with $D$, these increments are marginal even in the ideal case, and level off as $D$ increases~\cite{PhysRevLett.88.040404,DadaIJQI20011}\bluek{. Also,} the high sensitivity of the complicated measurement setup to errors will usually overwhelm these increments even for modest values of $D$, resulting in smaller violations than in the qubit case.  
Another advantage of our scheme 
 where an SPDC source is used is that non-maximal high-dimensional entanglement will not generally degrade the the verification of security. For example, \blueq{the spiral bandwidth~\cite{PhysRevA.68.050301} of the SPDC source will not generally degrade Bell violation, but will only limit the  effective number of OAM channels in the non-deterministic case}. 
Whereas, if generalised OAM-based Bell tests are used  without procrustrean filtering,
 then a small spiral bandwidth might cause a failure of the Bell test for an entangled state~\cite{Dada2011}.   

In summary, this paper has described a practical scheme in which a single CHSH-Bell test setup combined with a full projective measurement is sufficient for security verification even for a large-alphabet scheme capable of arbitrarily large key rates per run. The scheme is simpler to implement than existing generalizations of E91 protocol to high-dimensions because it circumvents measurements in mutually unbiased bases in high dimensions, while maintaining capacity for large key-rate and security against collective attacks. A second significant advantage is that non-maximal high-dimensional entanglement will not necessarily degrade \bluek{the verification} of security. We point out that the scheme is realisable using current technology by mentioning two examples for generating applicable source states, namely, spontaneous parametric downconversion and, more suitably, source of single pairs of entangled photons, such as semiconductor quantum dots. 
 From the point of view of real-world applications of high-dimensional QKD \blue{based on photon OAM in free space}, judicious selection of basis states~\cite{Pors:11,Malik:12} will increase resilience against decoherence induced by atmospheric turbulence in a free space \bluek{implementation}. Although this can be applied within the framework of this scheme, implementations with time bins~\cite{PhysRevLett.84.4737,PhysRevLett.93.180502} or path appear \blue{especially} promising for long distance applications. 
 The complexity of security verification in large-alphabet entanglement-based QKD makes it apparent that the simplified scheme presented here will likely enable otherwise infeasible secure key rates in 
 QKD, enabling more practical implementations of entanglement-based technologies. 

\section{Acknowledgements}
\bluek{The author acknowledges the Engineering and Physical Sciences Research Council  [EPSRC (grant numbers: EP/I023186/1, EP/K015338/1)] and the Scottish Universities Physics Alliance (SUPA) for funding, Prof. Brian Gerardot and Prof. Gerald Buller for support, and Dr. Ryan Warburton, Dr. Jonathan Leach and Prof. Miles Padgett for stimulating discussions}.


%


\end{document}